\documentclass{article}

\usepackage{arxiv}

\usepackage[utf8]{inputenc} 
\usepackage[T1]{fontenc}    
\usepackage{hyperref}       
\usepackage{url}            
\usepackage{booktabs}       
\usepackage{amsfonts}       
\usepackage{nicefrac}       
\usepackage{microtype}      
\usepackage{lipsum}
\usepackage{graphicx}
\usepackage{amsmath}
\usepackage{amssymb}
\usepackage{amsfonts}
\usepackage{amsbsy}
\usepackage{color}
\usepackage{subfigure}
\title{Square wave oscillation of  soliton in double-well potential trapped BEC}

\author{
 Decheng Ma \\
 Key Laboratory for Magnetism and Magnetic Materials of the Ministry of Education\\
 Lanzhou University\\
 Lanzhou 730000, China \\
   \And
  Chenglong Jia\thanks{email:cljia@lzu.edu.cn} \\
  Key Laboratory for Magnetism and Magnetic Materials of the Ministry of Education\\
  Lanzhou University\\
  Lanzhou 730000, China \\
}

\begin{document}
\maketitle

\begin{abstract}
We numerically investigate the soliton tunnelling process in double-well potential trapped Bose-Einstein condensate. Comparing with the usual low energy few particle tunnelling process, we find that the soliton tunnelling leads to massive particle transport between two wells. The {population imbalance} between two wells is not evolving sinusoidally with the time as the Josephson plasma oscillation, but shows a higher density contrast square-wave pattern due to the reflections at the trapping potential boundaries. Such unusual behavior clearly demonstrates the topologically stable, localized nature of solitons that propagate in a nonlinear medium without spreading. The square-wave oscillation of soliton also provides  measurable dynamics to define a qubit in cold altom system.
\end{abstract}


\section{Introduction}

Solitons are topologically stable fundamental excitations, which exist  in a variety of fields such as water in narrow channels, high-speed optical communication, molecular biology and astrophysics \cite{Dauxois2006}. Ever since the realization of atomic Bose-Einstein condensate (BEC), the study of solitons in this intrinsically nonlinear system has become an important topic both experimentally \cite{Burger1999,Denschlag2000,Anderson2001,Becker2008,Khaykovich2002,Strecker2002,Cornish2006}  and theoretically \cite{Jackson1998,Feder2000,Brand2002,Muryshev1999,Carr2001,Burger2002,Busch2000,Fedichev1999}. So far both dark solitons \cite{Burger1999,Denschlag2000,Anderson2001} and bright solitons \cite{Strecker2002,Cornish2006,Khaykovich2002} have been created in the BEC experiments, their dynamics and collisional behaviours indicate clearly the particle-like nature of solitons \cite{Becker2008,Carr2001,Burger2002,Busch2000,Fedichev1999}. On the other hand, the bosonic Josephson junctions have also been realized in BEC confined in a double-well potential 
\cite{Dalfovo1996,Milburn1997,Smerzi1997,Zapata1998,Raghavan1999,Giovanazzi2000,Zhang2001,Sakellari2002,Ananikian2006,Giovanazzi2008,Ichihara2008,Julia-Diaz2010,Jezek2013,Gati2006, Cataliotti2001,Albiez2005,Gati2007,Levy2007,LeBlanc2011,Trenkwalder2016}. The physics of Josephson effect is usually tackled by the two-mode model, where the two modes $\psi_{L,R}$ are usually taken either as the interacting ground state in each well \cite{Smerzi1997,Zapata1998,Giovanazzi2000}, or as a linear combination of the ground state and the first-excited state of the whole system \cite{Raghavan1999,Ananikian2006,Julia-Diaz2010,Sakellari2004,Gati2007,Adhikari2009}. 
But this crude approximation treatment can be inaccurate in many cases \cite{Leggett2001,Ananikian2006,Jezek2013,Valtolina2015,Japha2011}, especially for the strong interaction limit, to improve this method, some authors have adopted modified coupled-mode equations \cite{Ostrovskaya2000,Ananikian2006,Jia2008,Julia-Diaz2010}, leading to better prediction for stronger interaction and producing better agreement with numerical solutions of the
time-dependent Gross-Pitaevskii equation. There is also a different approach to treat the Josephson plasma oscillations by using the Bogoliubov approximation \cite{Burchianti2017,Dalfovo1999,Japha2011,Paraoanu2001,Meier2001}, which can give a better result for the Josephson plasma
oscillations even in strong interaction limits. However,  all these approaches are questionable to the tunnelling transport of  solitons \cite{Susanto2011}. 
Since solitons are topological-stable excitations, their existence in the BEC system leads to macroscopic quantum tunnelling and non-adiabatic evolution \cite{Sakellari2004}. The soliton related tunnelling dynamics also involve more high energetic modes \cite{Lee2007}, which is beyond the two-mode description and can also lead to the emergence of instabilities \cite{Middelkamp2010}. \\

To give a better understanding of the tunnelling dynamics of solitons in BEC,  we investigate the solitons' transport in one dimensional double-well potential 
by numerically solving the Gross-Pitaevskii equation. By comparing with the usual low energy tunnelling between the two potentials, we find that the soliton tunnelling leads to massive particle transport between two traps. The {population imbalance} does not oscillate sinusoidally with the time but show as a square-wave pattern, which is 
different from the Josephson oscillation predicated by the two-mode approximation. This kind peculiar behavior of soliton tunnelling process is due to the soliton back and forth reflection between the trapping potential, which as clear reveal of their particle-like character.

\section{Model}
\begin{figure}[htbp]
\centering
\includegraphics[height=5.0cm,width=8cm]{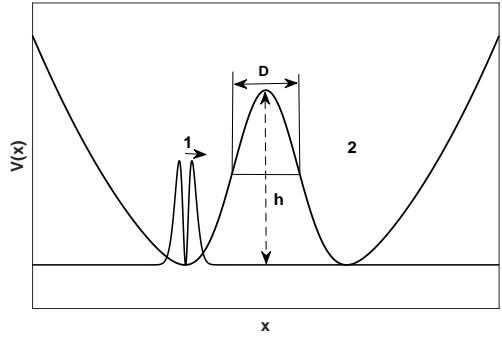}
\caption{Double well potential consisting of a harmonic potential and a Gaussian potential barrier in the middle. 
A right moving dark soliton is in the left well. 
}
\label{Fig::potential}
\end{figure}
Let us consider a system of $N$ bosons confined in a one-dimensional double-well potential $V(x)$ \cite{Sakellari2004a},  as show in Figure \ref{Fig::potential},
\begin{equation}
V(x)=\frac{1}{2}m\omega^2(x/s)^2+h e^{-(x/D)^2}
\end{equation}
where $m$ is the particle mass, $s$ is a dimensionless parameter to tune the width of the trapping harmonic potential. The potential barrier is placed in the middle of the potential, its height is given by the parameter $h$ and its width is specified by the parameter $D$. 
The dynamics of BECs in the double well can be described by the following Gross-Pitaevskii equation (GPE),
\begin{equation}
i\hbar \partial_t\psi(x,t)=[-\frac{\hbar^2\nabla_x^2}{2m}+V(x)+gN|\psi(x,t)|^2]\psi(x,t)
\end{equation}
Here $g$ is the inter-particle interaction strength. By choosing the oscillation length $a_0=\sqrt{\hbar/m\omega}$ as the length unit and the harmonic oscillation energy $E_{\omega}=\hbar\omega$ as the energy unit, we can rewrite the GPE in dimensionless form:
\begin{equation}
i\partial_{\tilde{t}}\tilde{\psi}=[-\frac{1}{2}\nabla_{\tilde{x}}^2+\frac{1}{2}(\tilde{x}/s)^2+\tilde{h} e^{-(\tilde{x}/\tilde{D})^2}+\beta|\tilde{\psi}^2|]\tilde{\psi}
\end{equation}
where $\tilde{\psi}=\sqrt{a_0}\psi$, $\tilde{t}=\omega t$, $\tilde{x}=x/a_0$, $\tilde{h}=h/(\hbar\omega)$, $\tilde{D}=D/a_0$  and $ \beta=gNa_0/(\hbar\omega)$. For simplicity, in the following we will omit the tilde.

\begin{figure}
\centering
\subfigure{\includegraphics[height=3.5cm,width=9cm]{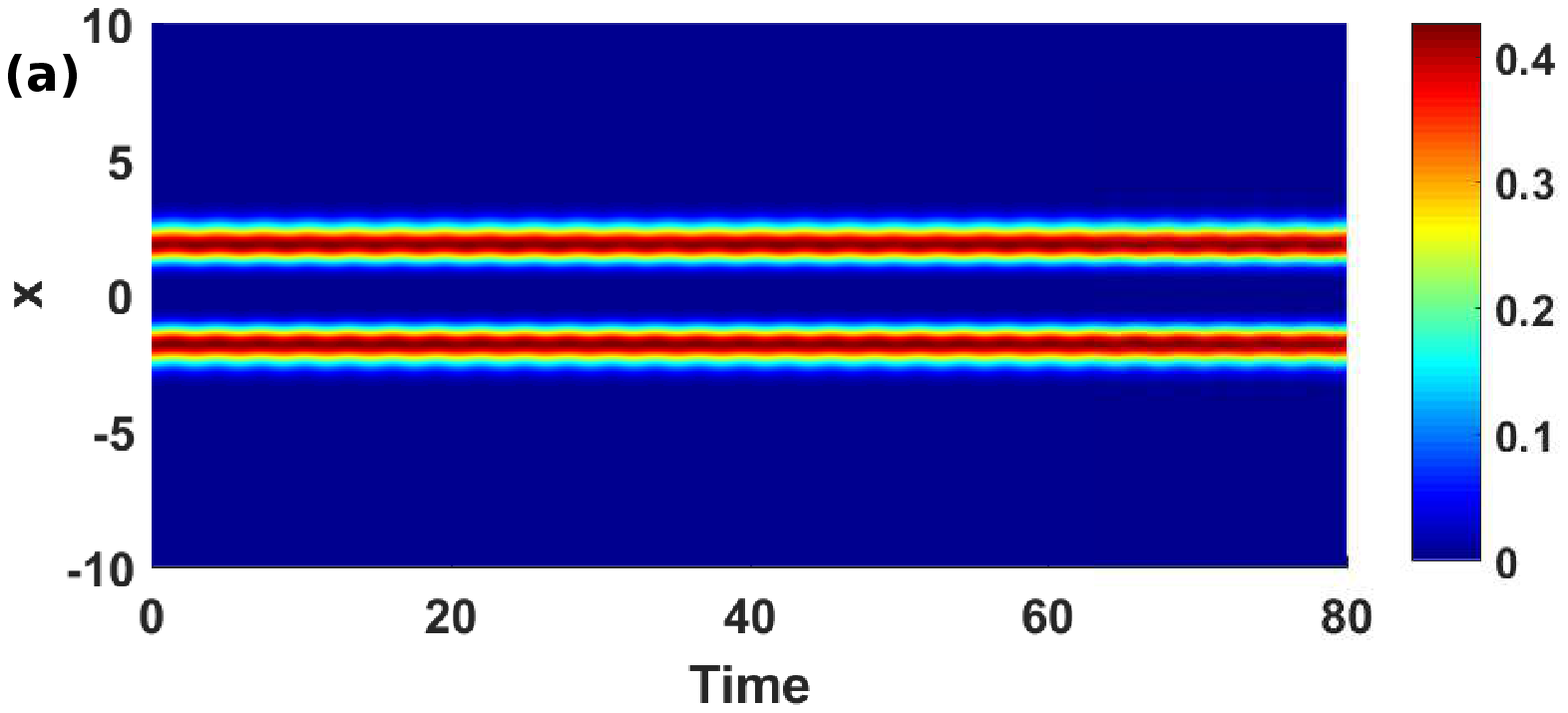}}
\subfigure{\includegraphics[height=3.5cm,width=9cm]{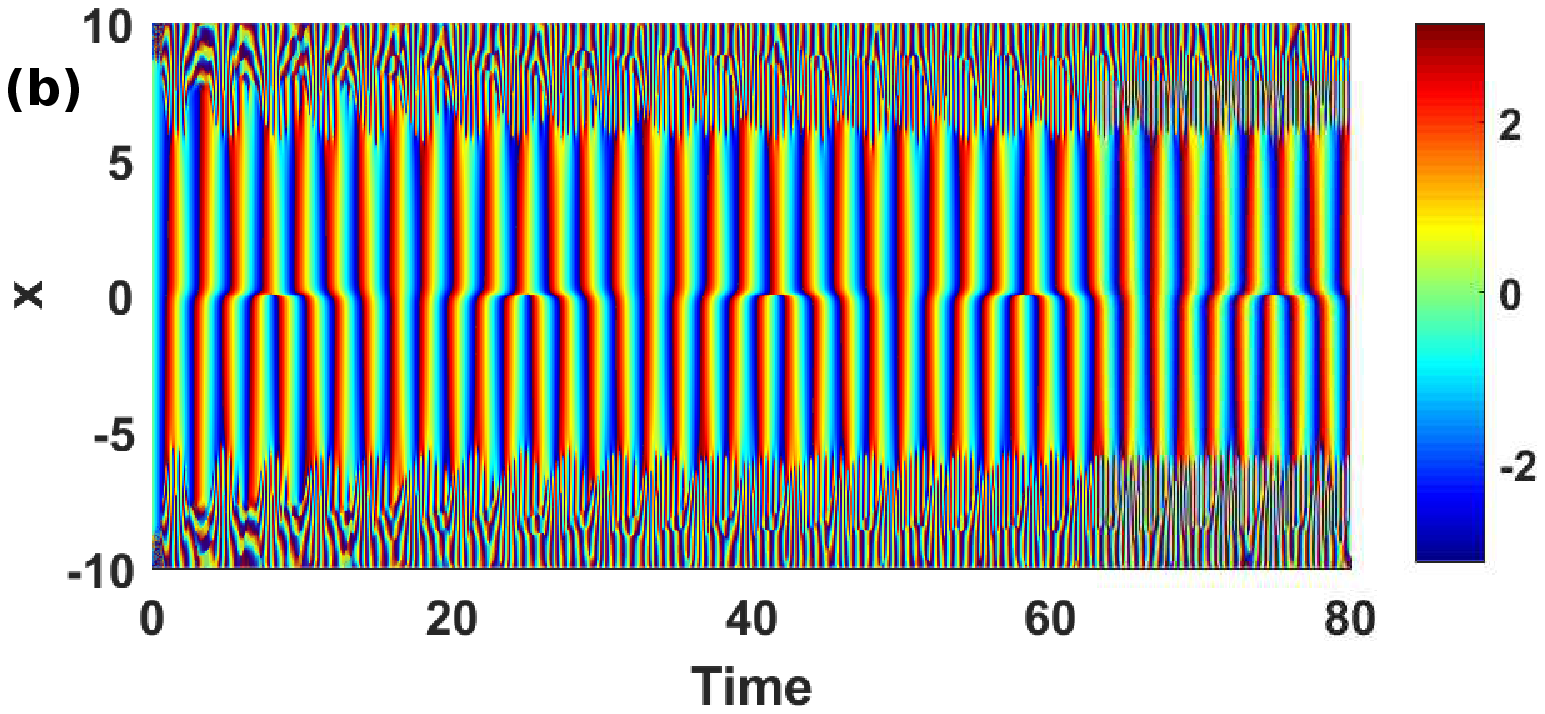}}
\subfigure{\includegraphics[height=3.5cm,width=7.5cm]{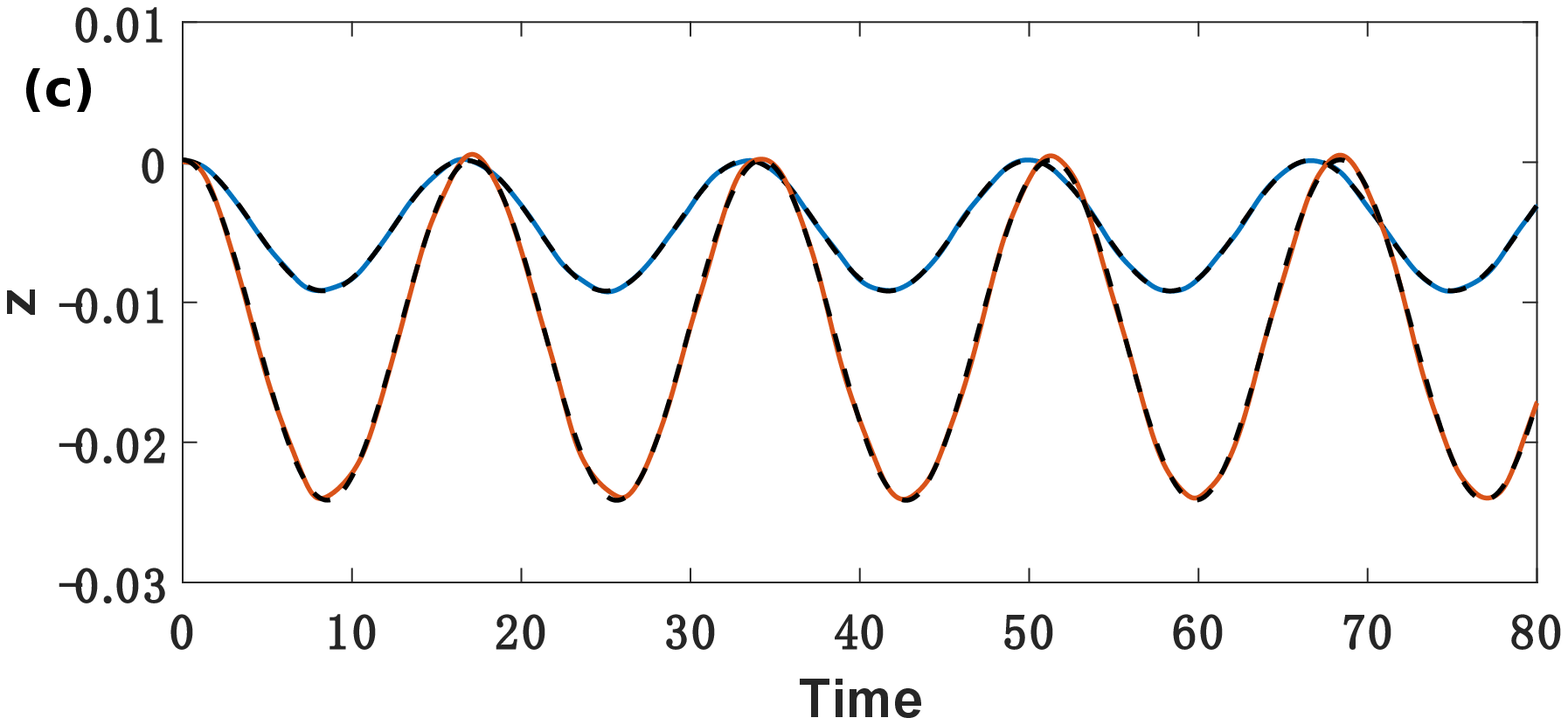}}

\caption{The dynamic evolution of the Josephson tunnelling in double-well trapped BEC :  (a) and (b) show  evolution of the coherent phase of the wavefunction  and the particle density in the absence of interaction ($\beta=0$),  (c) shows the population imbalance for $\beta=0$ (blue) and $\beta=5$ (orange) with the corresponding sine function fitting (black dash line)  . The rest parameters are $h=10, \,  s=1, \,  D=1$. }
\label{Fig::JC0}
\end{figure}

According to the two-mode model \cite{Smerzi1997}, we adopt a trial wavefunction
\begin{equation}
\psi = \psi_1(t)\Phi_1(x)+\psi_2(t)\Phi_2(x)
\end{equation}
where $\Phi_1$, $ \Phi_2$ are the {pseudo-ground state} solutions for each trap with the condition $\int\Phi_1\Phi_2\simeq0$ and $\int|\Phi_1|^2=\int|\Phi_2|^2=1$. Substituting the Eq.(4) into the GPE, we then have

\begin{equation}
\begin{split}
i\frac{\partial \psi_1}{\partial t}=(E_1^0+U_1N_1)\psi_1-K\psi_2, \\
i\frac{\partial \psi_2}{\partial t}=(E_2^0+U_2N_2)\psi_2-K\psi_1,
\end{split}
\end{equation}
with
\begin{gather*}
E^0_{1,2}=\int[\nabla\Phi_{1,2}|^2+\Phi^2_{1,2}V]d\mathbf{r},\\
U_{1,2}=\beta\int\Phi^4_{1,2}d\mathbf{r}, \\
K=-\int[(\nabla\Phi_1\nabla\Phi_2)+\Phi_1\Phi_2V]d\mathbf{r}.
\end{gather*}
Here we rewrite $\psi_{1,2}$ as $\psi_{1,2}=\sqrt{N_{1,2}}e^{i\theta_{1,2}}$, where $N_{1,2}$ and $\theta_{1,2}$ are the number of particles and phases in the trap 1, 2 respectively. It should be noted that $\psi_{1,2}$ is $x$-independent within the two-model approximation.
In terms of the phase difference $\phi=\theta_2-\theta_1$ and {population  imbalance} $-1<z=\frac{N_1-N_2}{N}<1$, the dynamic equations become

\begin{equation}
\begin{split}
&\dot{z}=-\sqrt{1-z^2}\sin\phi, \\
&\dot{\phi} =\Lambda z+\frac{z}{1-z^2}\cos\phi+\Delta E,
\end{split}
\end{equation}
where the time is scaled as $2Kt\rightarrow t$. $\Delta E$ and $\Lambda$ are defined as
\begin{gather*}
\Delta E=(E_1^0-E^0_2)/(2K)+(U_1-U_2)N/(4K),\\
\Lambda=(U_1+U_2)N/(4K).
\end{gather*}
For two symmetric traps, $E_1^0=E^0_2$ and $U_1=U_2=U$, we have $\Delta E=0$ and $\Lambda=UN/2K$. If the inter-particle interaction is negligible, which corresponding to $\Lambda\rightarrow 0$, Eq.(5)  yield Rabi-like oscillation in the population of each trap with frequency $\omega_R=2K$.

In Figure \ref{Fig::JC0} , we show the numeric simulation of  Josephson plasma
oscillations of Bose-Einstein condensate trapped by double-well potential $V(x)$ with parameters $h=10, \, D=1, \,  s=1$. In Figure \ref{Fig::JC0} (a) and (b), the inter-particle interaction strength is set to be 0, which is in the ideal Bose gas limit without any inter-particle interaction. While in Figure \ref{Fig::JC0} (c), we show the particle population imbalance under different inter-particle interaction strengths $\beta=0$ and $5$.
Given that the trial wavefunctions $\Phi_1$ and $\Phi_2$ are required to be orthogonal within the two-mode approximation, we prepare the initial state as the ground state $\psi_0(x)$ of a well-separated double well potential with high enough potential barrier to reduce the overlap of the wavefunction in each well. Then at time $t=0$, we translate the harmonic trapping potential by a small distance $\delta x=0.1$  to induce a little {imblance} between the two wells, which would lead to  the observable particle tunnelling between two wells \cite{Burchianti2017}.
  Based on the two-mode model, there should be a Rabi-like oscillation of density population in each trap. As we can see in the Figure \ref{Fig::JC0} (a), the trapped BEC are well separated, {while} their phases are coherent (cf. Figure \ref{Fig::JC0} (b)). The corresponding population imbalance in Figure \ref{Fig::JC0} (c) indeed shows smoothly sin-wave evolution, which is in good consistent with the two-mode approximation. 
 Comparing the population imbalance evolution of $\beta=0$ and $\beta=5$, we can see the presence of inter-particle interaction leads to an enhancement of the tunnelling amplitude and shift of periodicity.  \\

While for the case of solitons, given that they are topologically stable and preserve their form in long term due to the effects of non-linearity compensate for  dispersion \cite{book:17672}, we can treat soliton as a massive quasi-particle at the first step. 
For simplicity, we write the condensate wavefunction  in the form  $\psi(x,t)=\sqrt{\rho}e^{i\theta(x,t)}$, where $\rho$ is the density of condensate atoms and $\theta$  is the coherent phase of the condensate. According to the WKB approximation \cite{book:929512}, in the barrier region the wavefunction is in the form
\begin{equation}\label{soloton_t}
\psi(x,t) \simeq  \frac{C}{\sqrt{|p(x)|}}e^{-\int_{-a}^a dx|p(x)|}
\end{equation}
where $C=\sqrt{\rho\nabla\theta}$ and  $p=\sqrt{2m(V-E_{sol})}$. The barrier region is $[-a,a]$. $E_{sol}$ is the soliton energy given by $E_{sol}=\int_{sol} dx[|\frac{d\psi}{dx}|^2+\frac{1}{2}\beta(|\psi|^4-n_0^2)^2]$, in which the integral is integrated over the region where soliton extends \cite{book:17672} and $n_0$ is the background atom density. The corresponding tunnelling probability is approximately given by $T\simeq e^{-2\int_{-a}^a dx|p(x)|}$, {which can be obtained by the numerical calculations.}
Although Eq.(\ref{soloton_t}) can approximately  be used to describe single soliton tunnelling process, in reality due to the intrinsic nonlinear complexity of BEC system, the tunnelling dynamics can only be grasped with numeric simulation.\\

\section{Soliton tunnelling}

\begin{figure}
\centering
\subfigure{\includegraphics[height=3.5cm,width=9cm]{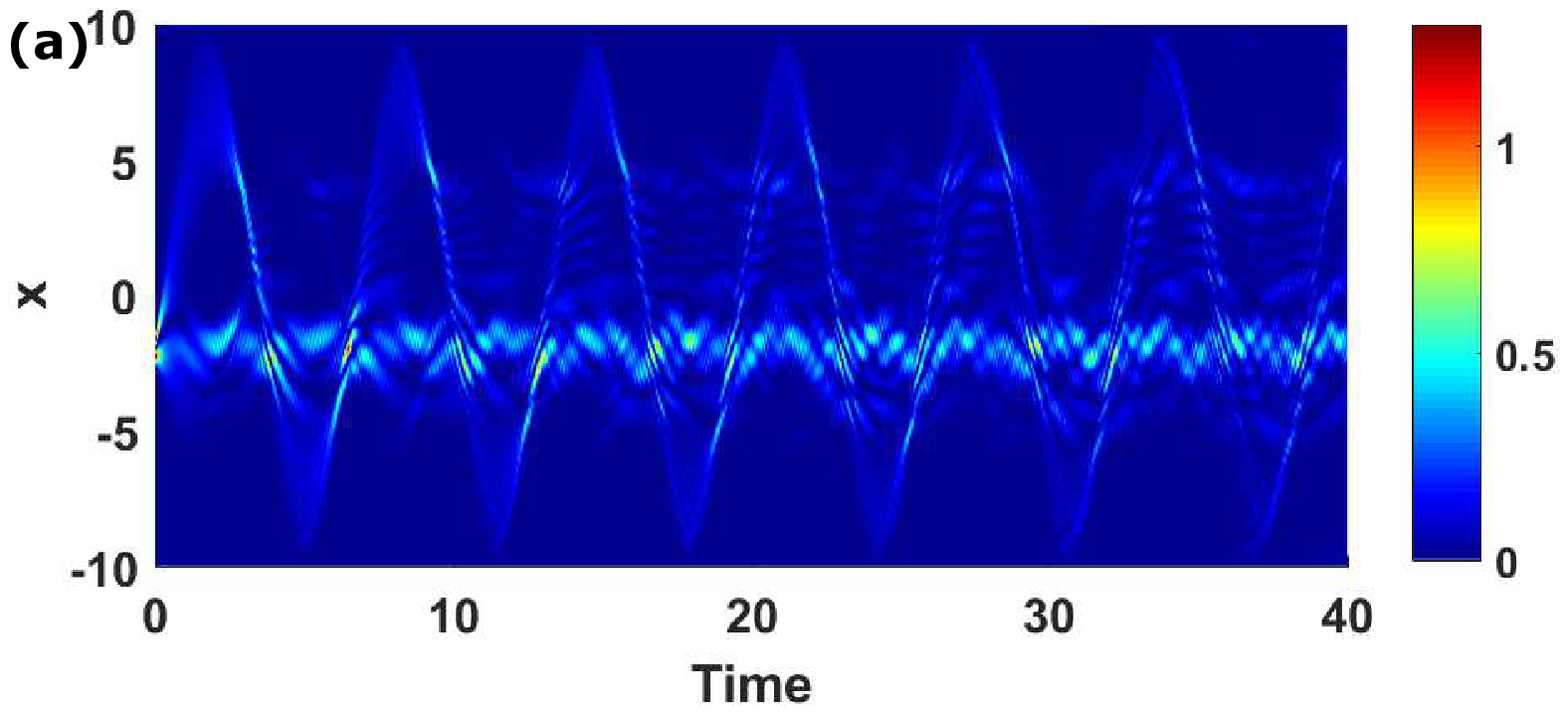}}
\subfigure{\includegraphics[height=3.5cm,width=9cm]{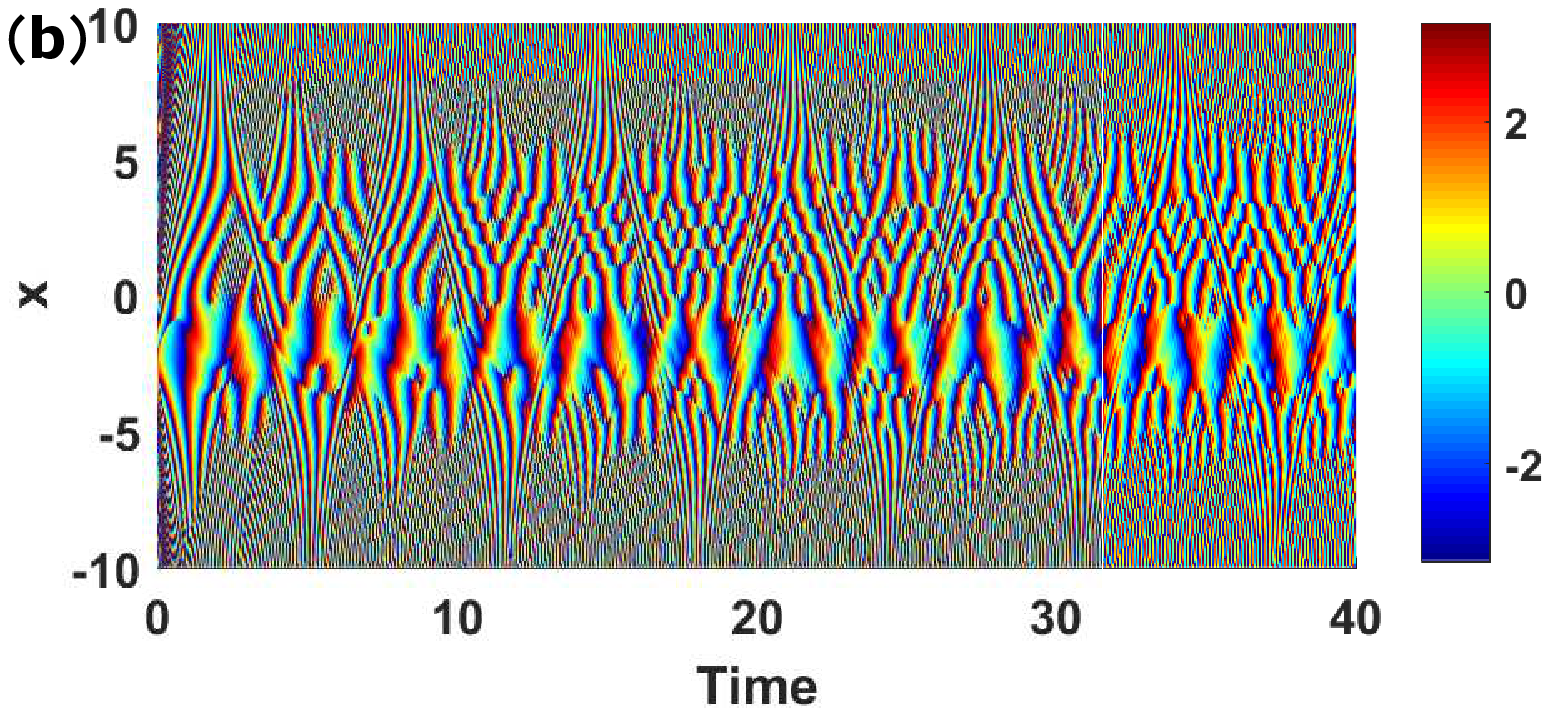}}
\subfigure{\includegraphics[height=3.5cm,width=7.5cm]{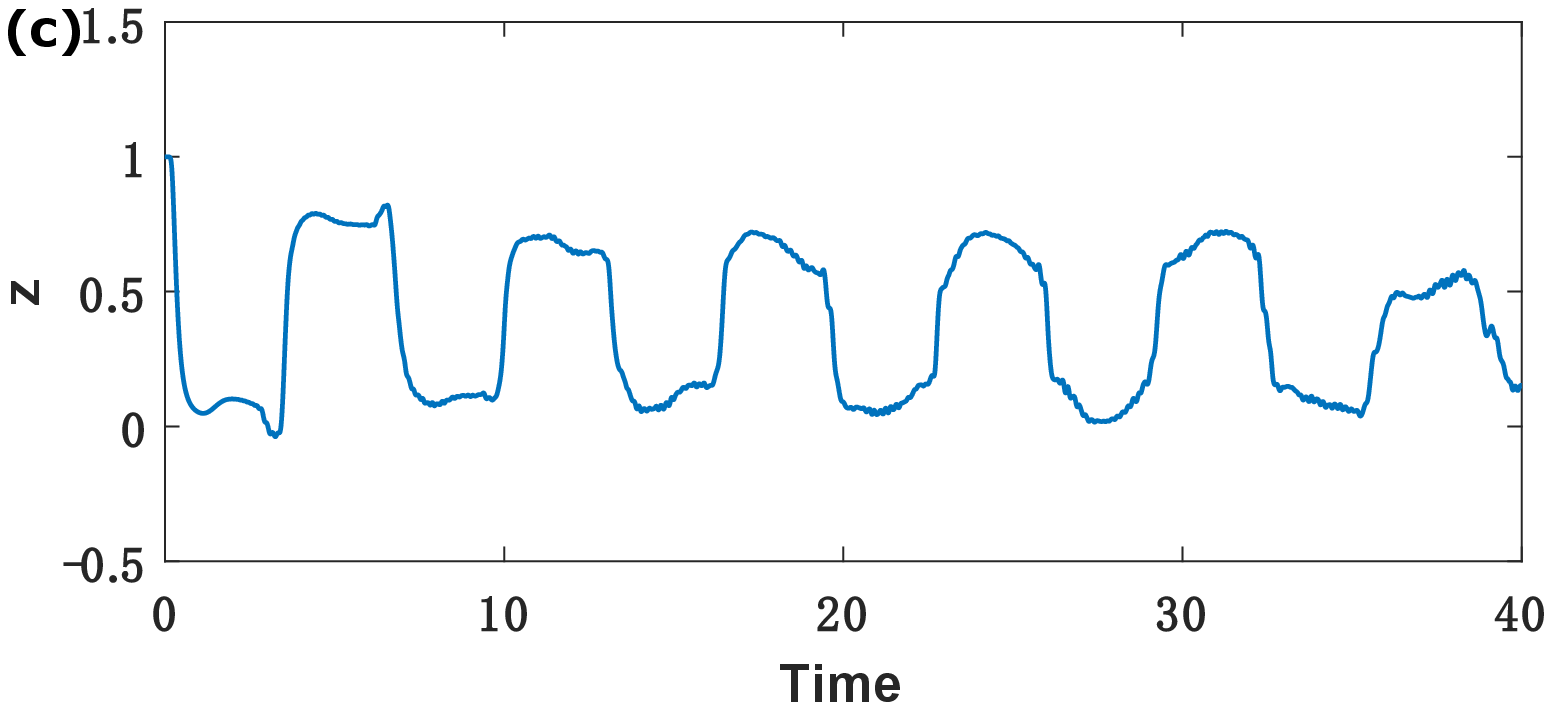}}
\caption{The dynamic evolution of the system with a single initial soliton {placed in the left well}: (a) the  particle density, (b) the corresponding phase, and (c) the population imbalance. The corresponding parameters are $h=10, s=1, D=1$, $\beta=5$ and the soliton is specified with $x_0=2$, $l=0.5$ and $c_1=5$. }
\label{Fig:soliton}
\end{figure}

In the numeric simulation, we using backward Euler pseudo-spectral scheme to computer the ground state of the GPE \cite{Antoine20142969}, then the dynamic evolution is obtained by applying the time splitting scheme \cite{Antoine201595}.

 We start the simulation for soliton tunnelling with a soliton localized in the left trap potential and set up the initial wavefunction as
\begin{equation}
 \psi_0(x)=A\tanh((x+x_0)/l)\cdot e^{-((x+x_0)^2/l)}\cdot e^{ic_1(x+x_0)/2}.
\end{equation}
where $A$ is the parameter that make the wavefunction  normalized to unity in the region $[-L,\, L]$, i.e., $A=\int_{-L}^{L}dx\tanh((x+x_0)/l)\cdot e^{-((x+x_0)^2/l)}$ with the parameter $l$ being used to vary the size of soliton. We have set the initial state with a uniform phase gradient $c_1/2$ to give the soliton a initial right-moving velocity.
Then we set the system to dynamic evolution with time-splitting scheme \cite{Antoine201595}.

\begin{figure}
\centering
\subfigure{\includegraphics[height=3.8cm,width=9cm]{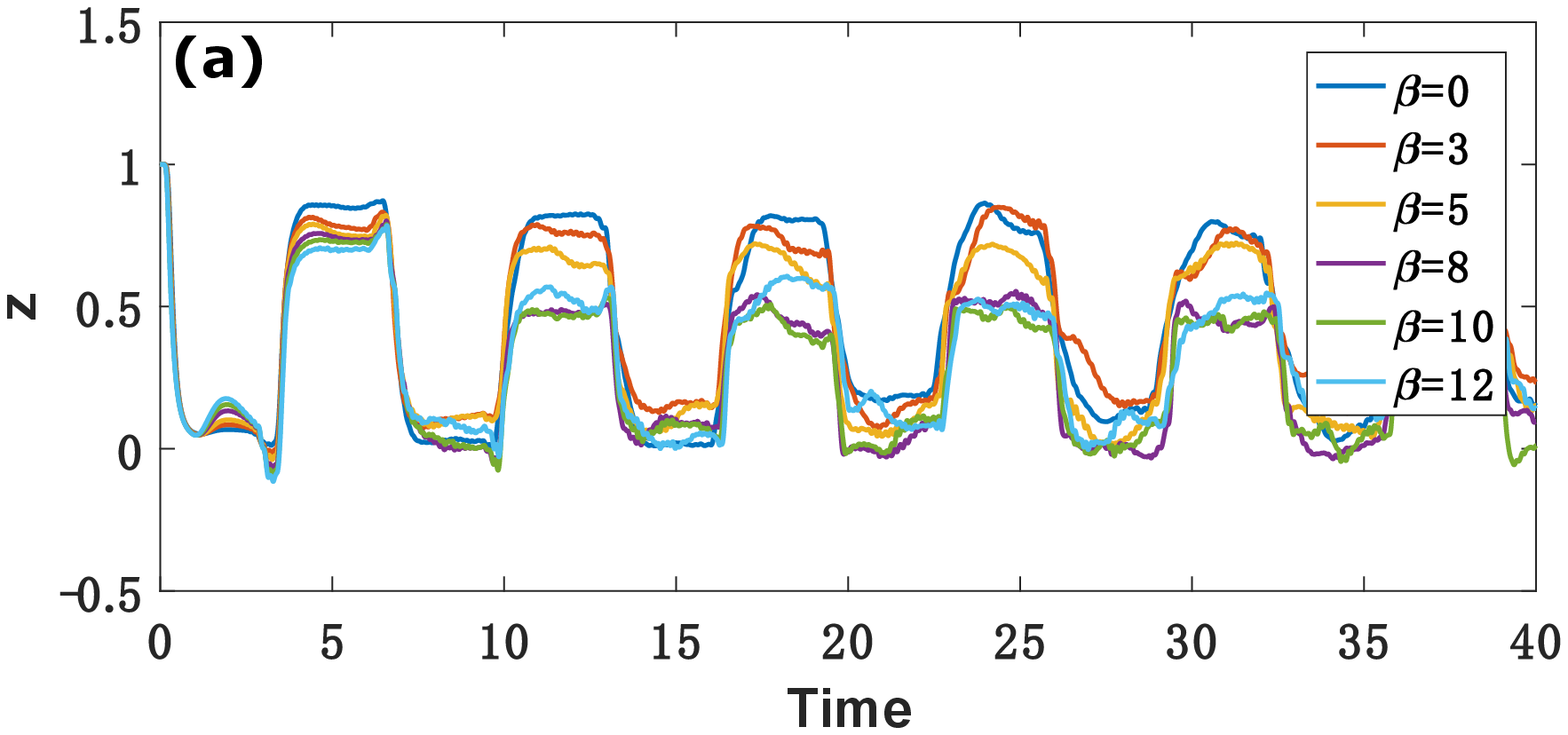}}
\subfigure{\includegraphics[height=3.8cm,width=9cm]{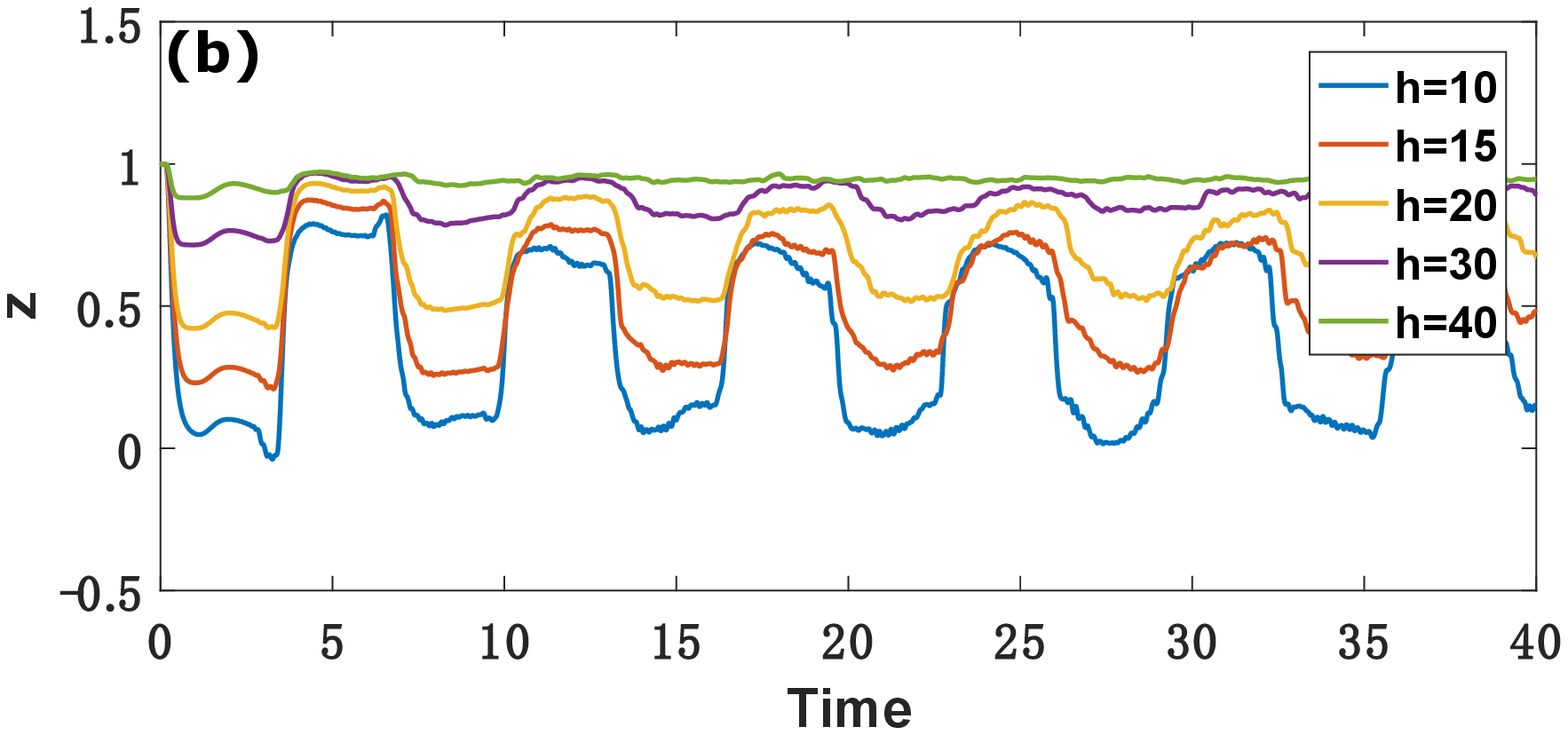}}
\caption{The population imbalance evolution for different interaction strength and different barrier height with soliton parameters $x_0=2$, $l=0.5$, $D=1$, $c_1=5$. (a) The population imbalance with potential barrier parameter $h=10$ and different interaction strength $\beta=0, \, 3, \, 5, \, 8, \, 10, \, 12$. (b) The population imbalance with interaction strength $\beta=5$ and different height barrier $h=10, \, 15, \, 20, \, 30, \, 40$. }
\label{Fig::h}
\end{figure}

In Figure \ref{Fig:soliton}, we show the soliton dynamic process in the case with parameter $h=10,\, s=1,\, D=1$ and $\beta=5$, while the soliton is specified with parameter $x_0=2, \, l=0.5$ and $c_1=5$. The particle density, the wavefunction phase, and the population imbalance are demonstrated respectively in  Figure  \ref{Fig:soliton}  (a), (b), and (c).  
Since the solitons are excitations composed by many bosons and stable against dispersion, in the transport process they can keep their shape and form in long term like a quasi-particle, which would give rise to massive bosonic tunnelling \cite{Sakellari2004}. Comparing with the  Josephson plasma oscillation case,
 as show in Figure \ref{Fig:soliton} (c),  the population imbalance gets a much more stronger signal and there are many sudden jumps in the population imbalance evolution. {Such peculiar pattern} is a clear signature of the soliton tunnelling. This massive phenomena is similar to the soliton oscillation \cite{Becker2008,Busch2000}, but in our case 
there are many solitons vanish and generated during the tunnelling process, which make it very hard to trace the dynamics of each soliton. 
Furthermore, we see from Figure \ref{Fig:soliton} (a), after the solitons tunnel through the potential barrier, they continue to move forward until reaching the potential boundary, then they are reflected and move backward. Such periodical reflection results in a square-wave like pattern in the evolution process, as show in the Figure \ref{Fig:soliton} (c).

\begin{figure}
\centering
\subfigure{\includegraphics[height=3.5cm,width=9cm]{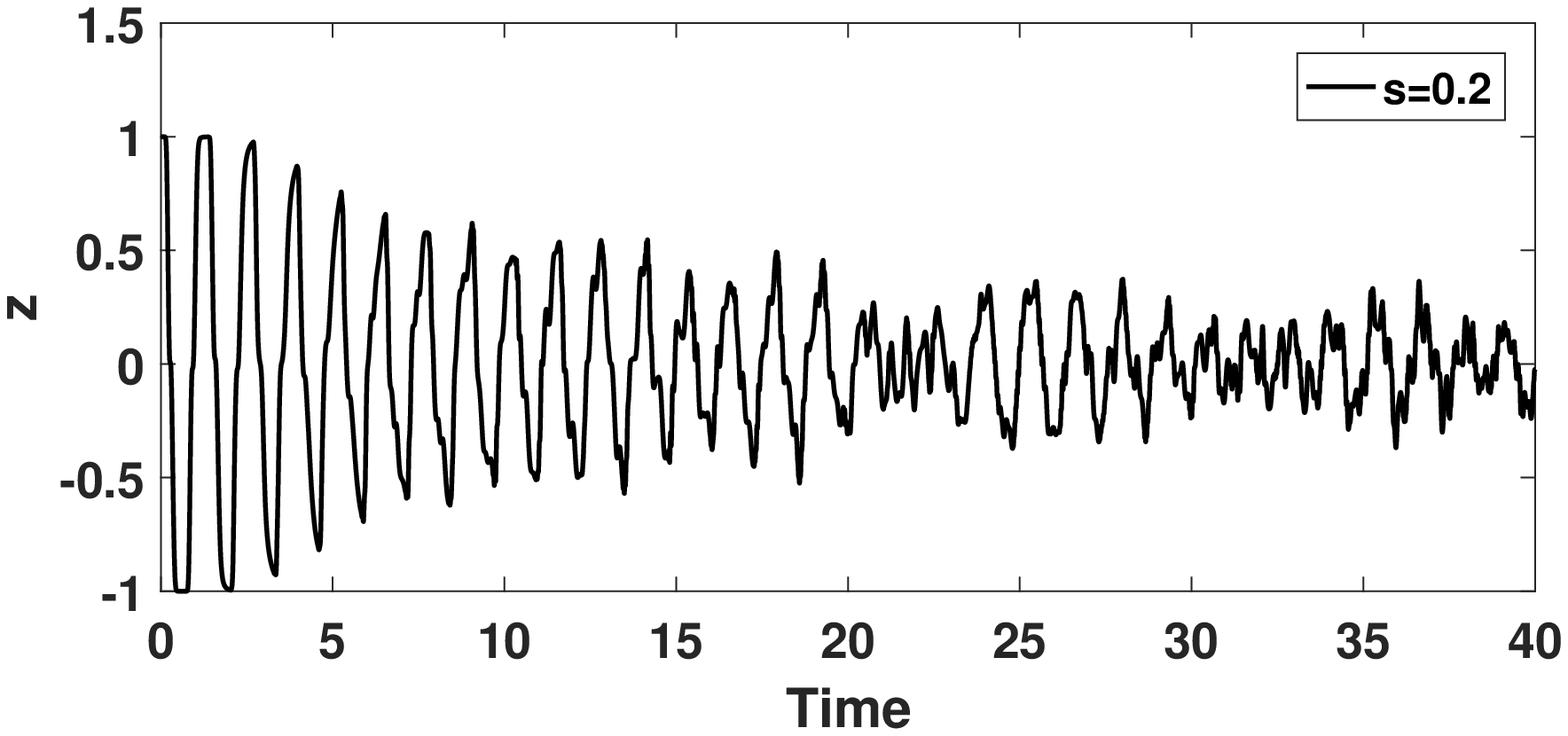}}
\subfigure{\includegraphics[height=3.5cm,width=9cm]{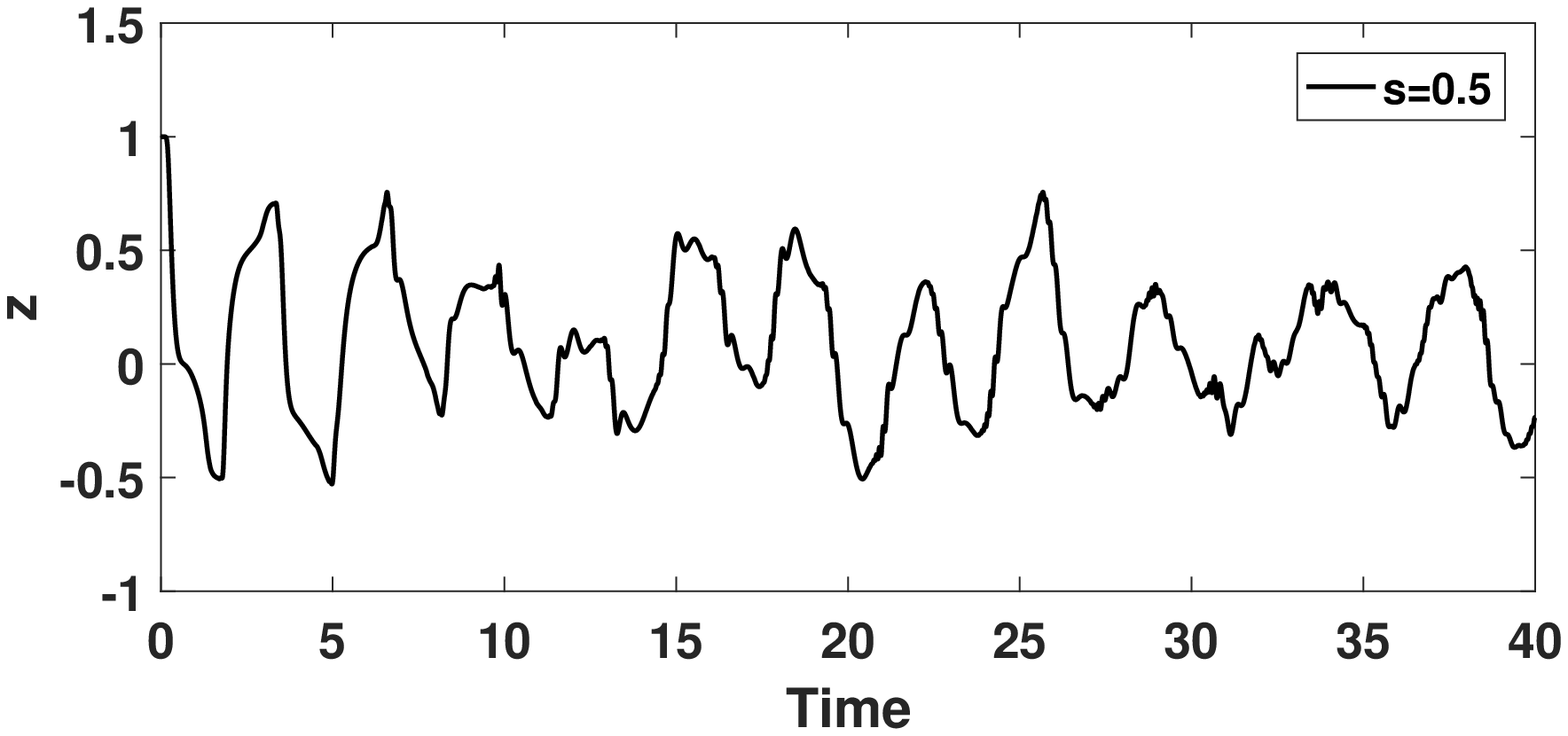}}
\subfigure{\includegraphics[height=3.5cm,width=9cm]{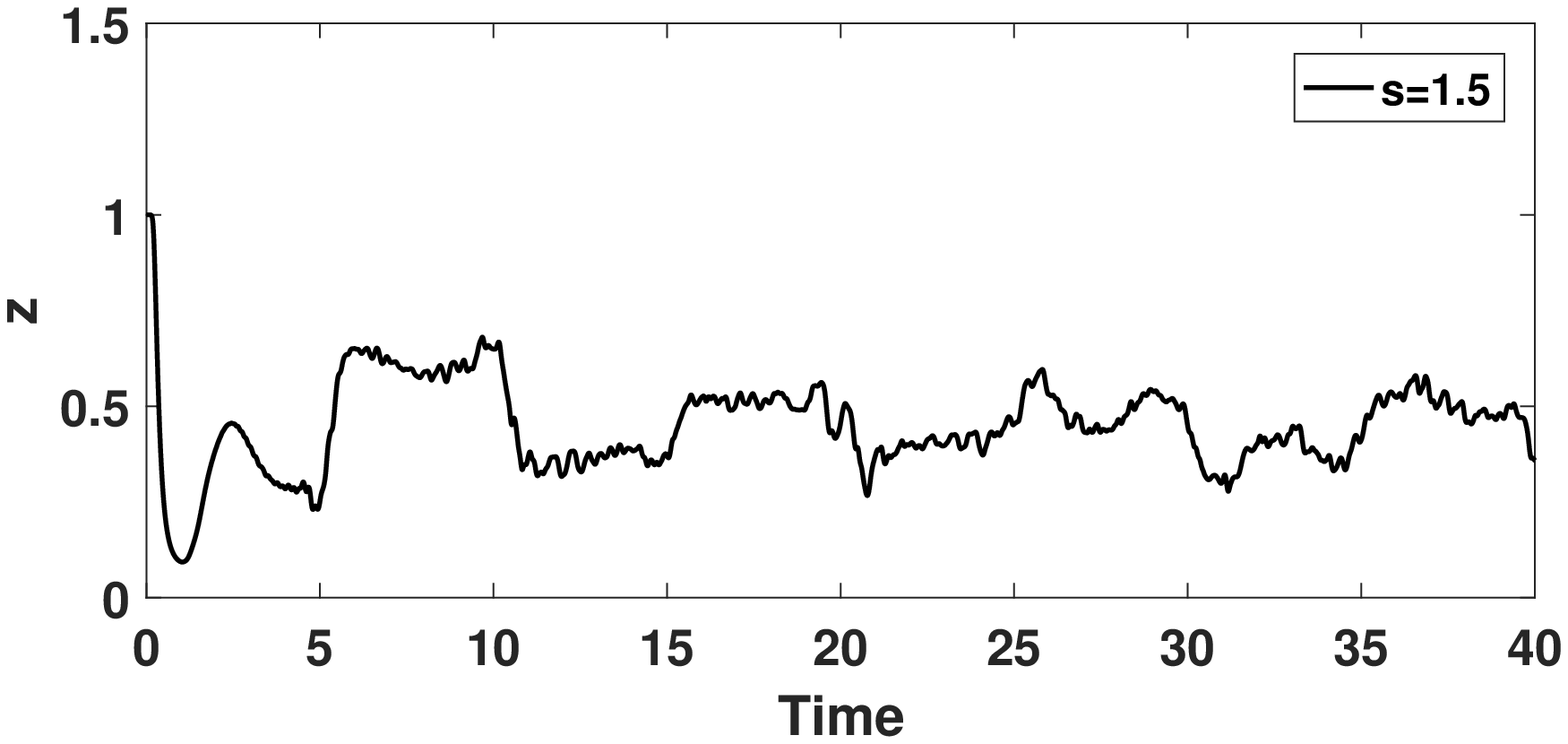}}
\caption{ The population imbalance as a function of time under different potential width  $s=0.2, \, 0.5$ and $1.5$.  Other parameters are $h=10, \, D=1, \, \beta=5$, $c_1=5.0, \,  x_0=2$ and $l=0.5$. }
\label{Fig::sv}

\end{figure}

\begin{figure}
\centering
\subfigure{\includegraphics[height=4cm,width=9cm]{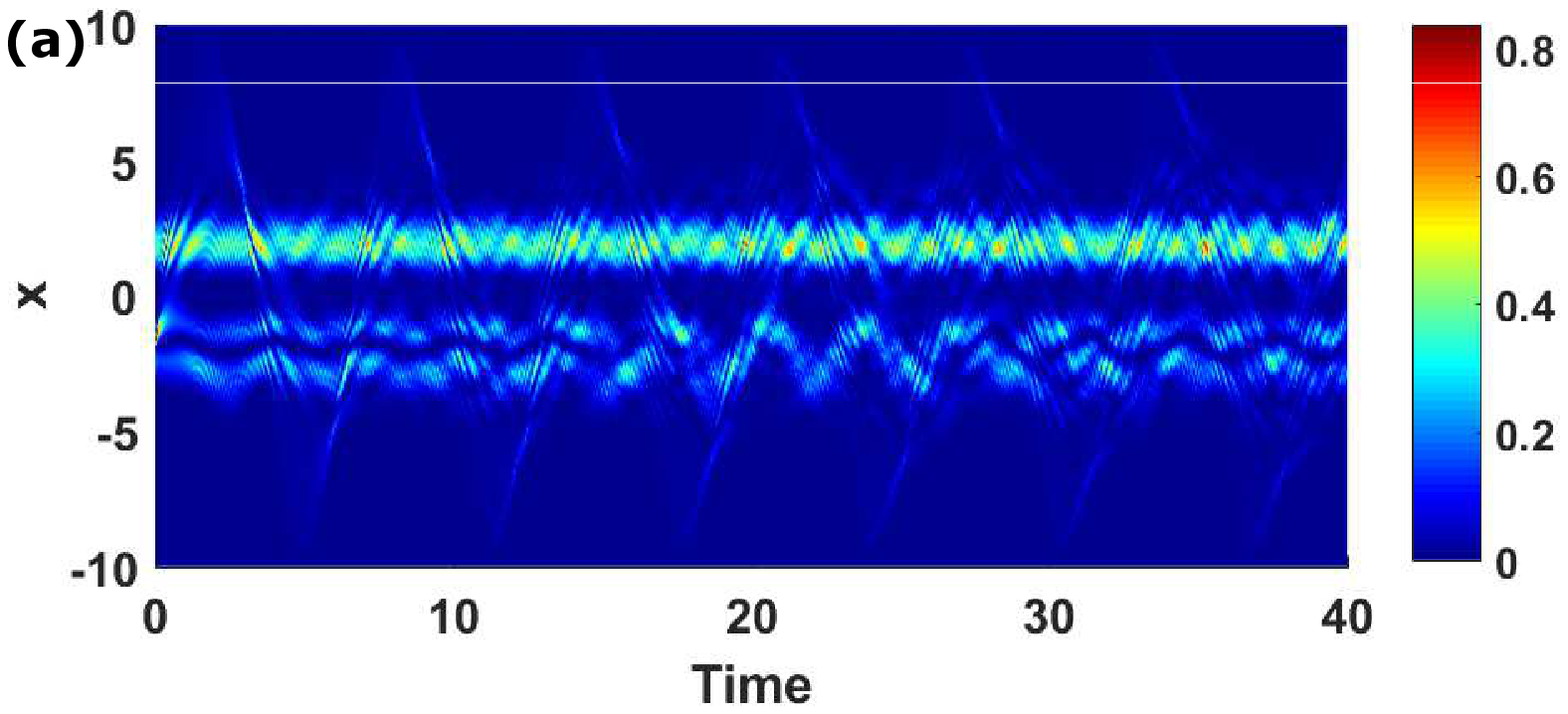}}
\subfigure{\includegraphics[height=4cm,width=9cm]{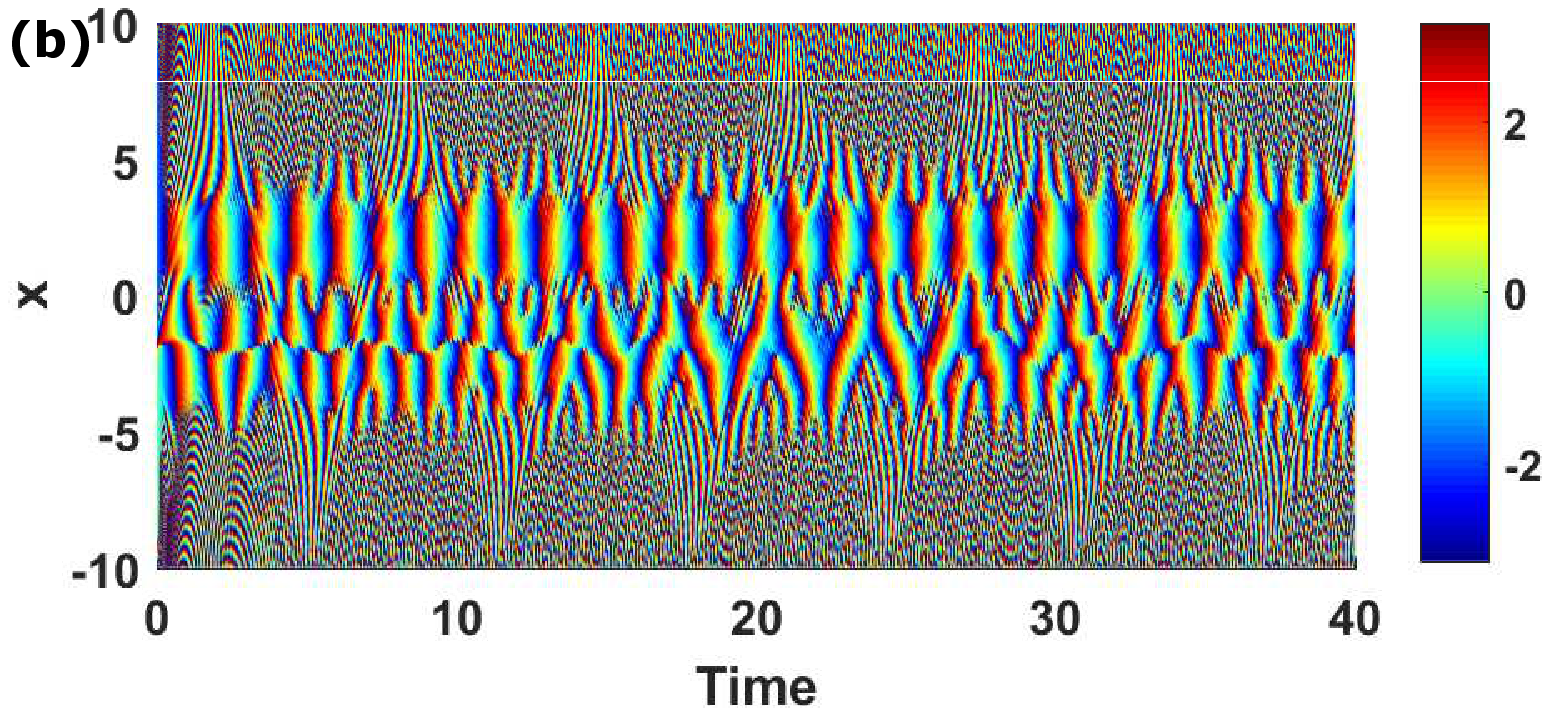}}
\subfigure{\includegraphics[height=3.5cm,width=7.5cm]{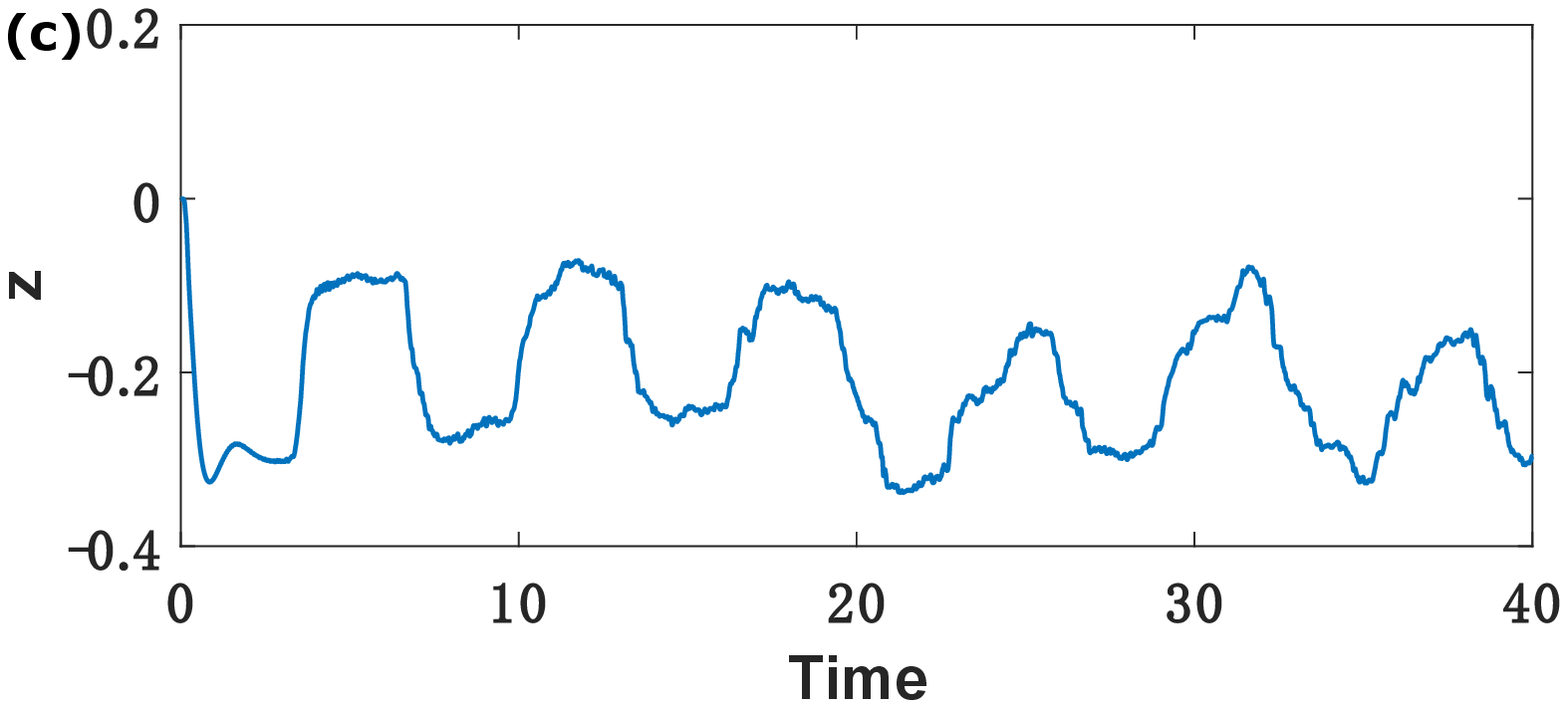}}
\caption{The dynamic evolution of (a) the particle density, (b) the corresponding phase, and (c) The population imbalance under phase imprinting with parameters $h=10, \, s=1, \, D=1 $ and $\beta=5$ and the phase imprinting parameters are $\alpha=\frac{3}{2}\pi, \, x_0=2.0, \, d=0.2$. }
\label{Fig::pi}
\end{figure}

To investigate to periodicity dependence, we vary the interaction strength parameter $\beta$ and the potential barrier height $h$ to see the change of  population imbalance evolution. In Figure \ref{Fig::h} (a), we vary the inter-particle interaction strength parameter to different values   $\beta=0, \, 3, \, 5, \, 8, \, 10, \, 12$, while keep the potential barrier parameter $h=10$ unchanged. Whereas, in Figure \ref{Fig::h} (b), we set the potential barrier height to different values  $h=10, \, 15, \, 20, \, 30, \, 40$ and keep the inter-particle interaction strength $\beta=5$ unchanged. The rest parameters are set to be  $x_0=2$, $l=0.5$, $D=1$ and $c_1=5$.  As we have expected, solitons possess an intrinsic quasi-particle nature: the interaction strength and the barrier height change the amplitude of population imbalance dramatically, but have little influence on the period.

Since the periodicity of the population imbalance is related to the solitons reflection at the potential boundaries, by changing the potential width parameter $s$ should lead to the change in the periodicity. In Figure \ref{Fig::sv}, we change the value of $s$ to $0.2, \, 0.5 $ and $1.5$, while retaining the rest parameters as $h=10, \, D=1, \, \beta=5$, setting the parameters that specified the initial soliton as $c_1=5.0, \, x_0=2$ and $l=0.5$. 
Just as expected, the narrower the potential is, the less time is needed for the solitons to bounce off between the potential. 
While the potential is narrower, the particles are trapped closer and the particle density increases, there would be more particles to participate in the soliton tunnelling process, which also lead to the enhancement of population imbalance amplitude. When the potential is broad, the periodicity increases and the amplitude also reduces.

Experimentally the mostly used method to create solitons in BEC is by applying phase-imprinting to the condensate \cite{Denschlag2000}. Numerically, we start our simulation with a initial state by applying a phase-imprinting, $\psi(x,0)=\psi_{0}(x)e^{i\theta(x)}$, where $\psi_{0}(x)$ is the ground state, and $\theta(x)=\alpha(1+\tanh((x-x_0)/d))$ is the corresponding phase to be imprinted onto the ground state wavefunction \cite{Denschlag2000}. The numerical results are shown in Figure \ref{Fig::pi} with the potential barrier height to $h=10$, the interaction strength $\beta=5$, and the phase imprinting $\theta(x)=\frac{3}{2}\pi(1+\tanh((x+2)/0.2))$. Since we have placed the phase onto wavefunction in the left well, the solitons are firstly generated in the left trap with an initial velocity $v=\nabla_x\theta$ to travel to the right. 
The phenomenon is similar to former discussed case with initial state of a single soliton. However, one should note that the velocity of soliton is now position-dependent due to the inhomogeneous phase gradient in space. 
Therefore, in the long run {the square-wave-like pattern} of population imbalance can not retain and the quasi-periodicity will also change. With phase-imprinting method  similar in \cite{Susanto2011}, we have created the solitons in our system, but there is no Josephson tunnelling of dark solitons in real space.\\

\section{Conclusion}

In summary, by numeric solving the GPE, we find that {the population imbalance due to soliton tunnelling} evolves with a quasi-periodic square-wave pattern, which is different from the  Josephson oscillation predicated by the two-mode approximation. Its periodicity is determined by the trapping potential width, while the interaction strength and potential barrier height only lead to the  change of population imbalance amplitude and have little influence on the period. Furthermore, the soliton tunnelling in BEC  gives rise to much  stronger dynamic signal than  Josephson effect, which  makes it much more robust against fluctuations and perturbations.  
Such peculiar square wave pattern and robust character of soliton tunneling can be served as a good candidate in the realization of qubit for quantum information based on the collective properties of BEC \cite{Shaukat2017, Strauch2008}.

\section{Acknowledgment}

This work is supported by the National Natural Science Foundation of China (No. 11474138), the Program for Changjiang Scholars and Innovative Research Team in University (No. IRT-16R35), and the Fundamental Research Funds for the Central Universities. Decheng Ma thank Romain Duboscqa for his helps on the usage of GPELab.

\section*{References}
\bibliographystyle{unsrt}
\bibliography{reference}

\end{document}